\documentclass{article}
\usepackage{amsfonts}
\usepackage{amssymb}  
\usepackage{amsmath}
\usepackage{epsfig}

\def\<{\left\langle}
\def\>{\right\rangle}

\begin{document}

\title{\hspace{4.1in}{
} 
\\
\vspace*{1cm} \bf Supersymmetry breaking in ISS coupled to gravity}
\author{\vspace*{0.5cm} \bf Z. Lalak, O.~J.~Eyton-Williams
 \\
 Institute of Theoretical Physics, University of Warsaw, 
       00-681 Warsaw, Poland\\}
\date{}
\maketitle
\centerline{} 
\begin{abstract}
  \noindent We analyse the breakdown of supersymmetry in an ISS model
  in the presence of gravity, under the requirement that the
  cosmological constant vanishes dynamically.  The gravitational
  backreaction is calculated in the metastable minimum and, in
  conjuction with the condition $V=0$, this is shown to generate
  non-zero F-terms for the squarks.  Once the squarks are coupled to
  the messenger sector, a gauge mediation scheme is realised and it
  leads to a distinctive soft spectrum, with a two order of magnitude
  split between the gaugino and the soft scalar masses.
\end{abstract} 

\vskip 1cm

\section{Introduction}\label{sec:Intro}
In this letter we analyse the meta-stable point of a simple Intriligator,
Seiberg and Shih (ISS) \cite{Intriligator:2006dd} model, within the
framework of supergravity.  This allows us to cancel the cosmological
constant, which we opt to do by the simplest possible method: adding a
constant, $W_0$, to the superpotential.  This is sufficient to
generate a physically reasonable gravitino mass and balance the new
negative contribution to the potential against the original positive contribution
coming from the ISS potential.

We recompute the one-loop effective potential in supergravity and use
this to compute the gravitational backreaction on the global vacuum
\footnote{For an earlier study with somewhat different findings, see
  \cite{Kitano:2006wz}.  An interesting study that coupled a realistic
  moduli sector to an ISS model can be found in
  \cite{Serone:2007sv}.}.  The perturbations are shown to be small, as
one expects from gravitational corrections, but non-trivial. We stress
that it is necessary to consider gravitational corrections, even
though we know they are small.  They are important when determining
the expectation values of the fields, most notably the moduli,
but the remaining fields are shifted more than dimesional analysis
would suggest. The most interesting effect is the
generation of non-zero, but Planck suppressed, F-terms for the
magnetic quarks. Hence, there appear two distinct scales in
the sector that breaks supersymmetry.

It is interesting to calculate the relative importance of several
mediation mechanisms in this setup, specifically anomaly, gravity and
gauge mediation.  We give order of magnitude estimates for the soft
masses spectrum generated by these three mechanisms and argue that the
spectrum can have a striking gap between the gaugino masses and the
soft scalar masses.  This is reminiscent of split SUSY
\cite{ArkaniHamed:2004fb}, but the split is not allowed to be
arbitrarily large since it is constrained by the requirement that
$V=0$ in the meta-stable minimum.

We note that setting $V=0$ at tree-level is clearly not sufficient to
guarantee it remains close to zero when loop corrections are included.
As discussed in detail in \cite{deAlwis:2006nm, deAlwis:2007qx} one
generically expects both the logarithmic corrections present in the
SUSY theory (albeit with gravitationally corrected masses) and
quadratically divergent contributions, $V_{\mbox{quad.}}=
\frac{1}{32\pi^2}\mbox{STr}\mathcal{M}^2 \Lambda^2$, to be present if
the theory is cut-off at $\Lambda$.  However, as noted in
\cite{deAlwis:2007qx} and discussed further in \cite{Ferrara:1994kg},
this contribution is determined by the geometry of the K\"ahler
potential and the number of degrees of freedom in the effective
theory, and in principle it is possible for it to vanish.  Even if it
remains, its presence is not necessarily particularly damaging, since
it is fixed by the size of $m_{3/2}$.

The potential can then be parametrised as (with $M_P$ set to $1$ and
$V_{\mbox{log}}$ denoting the logarithmic one-loop contribution):
\begin{equation}
  V= V_F+V_{\mbox{log}}+(Z-3) m_{3/2}^2
\end{equation}
where $Z$ is a parameter encapsulating our ignorance about UV effects
and is $\mathcal{O}(\frac{1}{32\pi^2})$ to
$\mathcal{O}(\frac{N_{\mbox{\tiny TOT}}}{32\pi^2})$, where
$N_{\mbox{\tiny TOT}}$ is the total number of chiral fields.  If
$Z<0$, the condition $V=0$ is satisfied by a smaller $W_0$ than is
required to cancel the tree-level potential.  Since we know that
$|Z|\propto \Lambda^2$ this implies that it must be possible to chose
a cut-off small enough that $W_0$ will not change dramatically, and
our results will be qualitatively unchanged with respect to the case
with the quadratically divergent term omitted.  We have assumed, and
prove in appendix~\ref{sec:appB}, that the derivatives of
$V_{\mbox{quad.}}$ are similar in form and magnitude to
$V_{\mbox{tree}}$'s.  It is interesting to note that
$V_{\mbox{quad.}}$ can play an important role in the potential despite
being generated by gravity, in close analogy to the role played by $-3
e^K |W|^2$.  This is in contrast to the gravitational corrections to
the logarithmic potential, which are negligible in comparison to the
globally supersymmetric terms.  Naturally, we still have to re-tune to
get $V=0$, but the loop corrections do not increase the degree of
tuning required.  Finally, if $Z\gtrsim 3$ it is clear that these
models break down and the cosmological constant cannot be tuned to
zero.  This will not be the case unless the cut-off is close to the
Planck scale.  In these models, the relevant cut-off is the scale at
which the magnetic description is no longer valid as supersymmetry is
best described by the low energy variables in the magnetic theory.

We have implicitly assumed, in using $N=1$ supergravity formalism that
the UV preserves one supersymmetry.  On top of this, for simplicity's
sake, we assume that the sole source of SUSY breaking is the ISS
sector, with the constant $W_0$ setting the scale of $m_{3/2}$,
essentially postulating that $W=W_{\mbox{ISS}}+W_{\mbox{UV}}$,
$\<W_{\mbox{UV}}\>=W_0\neq 0$, that $F_{\mbox{UV}}\ll F_{\Phi}$ and
that the UV has been decoupled.  While the constant can be dynamically
generated in a explicit model we do not attempt to do so here (for an
example where a KKLT model \cite{Kachru:2003aw} is used in the UV see
\cite{Abe:2006xp}).  Finally we note that the UV's contributions to
the K\"ahler geometry, and hence $V_{\mbox{quad.}}$, are uncalculable,
but should be small on dimensional grounds.

\section{Global ISS review}
ISS showed that meta-stable SUSY breaking is possible in a wide class
of remarkably simple models.  One of their main examples is
supersymmetric QCD with $N_f$ flavours and $N_c$ colours.  If one lies
in the free magnetic range, $N_c<N_f<\frac32 N_c$, then the low energy
theory is strongly coupled, but admits a dual interpretation in terms
of IR-free, magnetic variables.

The tree-level potential in the magnetic theory is given by an
R-symmetry preserving O'Raifeartaigh model \cite{O'Raifeartaigh:1975pr} and so SUSY has to be
spontaneously broken: $F_i=0$ cannot be satisfied for all fields.

The tree-level superpotential in the magnetic theory is given by:
\begin{align}
  W_{\mbox{tree}}=h\mbox{Tr}\bigl(\phi \Phi \tilde{\phi}\bigr) - h \mu^2
  \mbox{Tr}(\Phi)
\end{align}
where $\Phi$ transforms as $N_f \times \overline{N_f}$, $\phi$:
$(\overline{N_f},N)$, $\tilde{\phi}$: $(N_f,\overline{N})$,
$N=N_f-N_c$, the number of squark flavours in the magnetic theory and we
denote the parts of $\Phi$ that will later obtain expectation values
as follows: $\Phi=\left(\begin{array}{cc}
    \Phi_1 & 0 \\ 
   0 & \Phi_0\end{array} \right)$.   The K\"ahler potential is canonical. 

Considering the tree-level superpotential in isolation one finds that
the lowest energy state is a moduli space parametrised by

\begin{equation}
  \Phi=\left( \begin{array}{cc}
      0 & 0 \\
      0 & \Phi_0\end{array} \right), \quad \quad \phi= \left(\begin{array}{c} \phi_0 \\ 0 \end{array} \right), \quad \quad \tilde{\phi}^{T}= \left(\begin{array}{c} \tilde{\phi}_0 \\ 0 \end{array} \right), \quad \quad \phi_0 \tilde{\phi}_0 = \mu^2 \, \mathbb{I}_{N_c \times N_c}. 
\end{equation}

Since SUSY has to be broken, the potential is positive definite and is found to
have an expectation value of $V=h^2 N_c \mu^4$.  When the one-loop effects
are included the moduli space is lifted and, aside from flat
directions identified with Goldstone bosons, a unique minimum is found
at:
\begin{equation} \Phi=0, \quad \quad\phi_0=\tilde{\phi_0}=\mu \,
  \mathbb{I}_{N_c \times N_c}. \end{equation}

In addition one must include the non-perturbative, R-symmetry
violating contribution:
\begin{align} \label{nnpert}
  W = Nh^{N_f/N}\bigl(\Lambda_m^{-(N_f-3N)}\det(\Phi)\bigr)^{1/N}.
\end{align}
Notice that the exponent of $\Lambda_m$,
$-\left(N_f-3N\right)=-(3N_c-2N_f)$, is always negative in the free
magnetic range.  Hence the coefficient of the determinant grows as the
cut-off shrinks.

Since the non-perturbative piece is R-symmetry violating a SUSY preserving
minimum must exist  \cite{Nelson:1993nf}, created by the non-perturbative piece. In global
SUSY \footnote{The situation could be improved in Sugra if the SUSY
  preserving point also had $W=0$ and the SUSY breaking point $V=0$,
  but this is difficult to obtain, and not the case here.  In-fact,
  the difference in the energy density is increased by the negative
  contributions from $W\neq0$.} this must be at a lower energy than
the SUSY breaking minimum.

\subsection{A note on dynamical scales}

We now calculate
the relationship between the dynamical scales, $\Lambda$ and $\Lambda_m$, of the electric and magnetic theories, respectively. We make use of the
relevant part of the dictionary given in ISS's paper and the duality
relation, given by:
\begin{align}
h= \frac{\sqrt{\alpha} \Lambda}{\widehat{\Lambda}}
\end{align}
and
\begin{align}
\Lambda^{3N_c-N_f}\Lambda_m^{2N_f-3N_c}=(-1)^{N_f-N_c}\widehat{\Lambda}^{N_f}.
\end{align}
where $\hat{\Lambda}$ is a dimensional parameter in the magnetic theory,
related to the electric quark mass, $m_0$, and the magnetic quark mass, $
\mu$ through the following relation: $\hat{\Lambda}=-\frac{\mu^2}{m_0}$

If we assume that the order one number, $\alpha$, appearing
in the K\"ahler potential for the electric mesons
\footnote{$K_M=\frac{1}{\alpha |\Lambda|^2} \mbox{Tr} M^\dagger M$.} is
simply 1 and that $h=1$, then it follows that all three scales,
$\Lambda$, $\Lambda_m$ and $\hat{\Lambda}$ are identified, up to
flavour dependent phases.  Above this scale the electric description
is valid, while the magnetic description is valid below.

\section{Locally supersymmetric  ISS}\label{sec:Local ISS}
If one simply promotes ISS to having a local supersymmetry without
including any additional physics, the results are not significantly
perturbed near the minima of the SUSY theory.  

However, the picture changes if any other terms appear in the
superpotential. Any new physics that generates a non-zero $\<W\>$,
necessary to have a finite gravitino mass and cancel the cosmological
constant, will at least interact gravitationally with the moduli.

Even the simplest possible modification, the addition of a small ($\ll 1$)
constant, $W_0$, to the superpotential, is sufficient to push the
pseudo-moduli to large expectation values.  This is not altogether
surprising since the global, tree-level potential is independent of
the pseudo-moduli and so their entire potential is given by Planck suppressed,
non-renormalisable operators once supergravity corrections are included.  As such,
the natural scale for their expectation values is $M_p$.

However, one-loop effects should not be ignored.  In the vicinity of
the metastable point the logarithmic one-loop potential generates mass
corrections of order $h^2\mu$ multiplied by a loop suppression factor.
For comparison, the typical contribution to the logarithmic
part of the one-loop potential from the gravitational effects
is $(h^2\mu^3/M_{P})^{1/2}$ \footnote{This can be derived assuming
  that $W_0\sim \mu^2$ which will be required for cancellation of the
  cosmological constant; $h K_i W_0 F$ gives a contribution to the
  mass square matrix of order $h^2 K_i W_0$, i.e.  $h^2 \mu^3$.}.
Hence, the gravitational corrections to the logarithmic one-loop potential, while non-zero, will be small. This does not hold for the
quadratic corrections which give mass corrections of order $h\mu$,
suppressed by the cut-off and a loop factor. 

In the following section we will consider the following simplified model, with the
non-perturbative piece removed.  This will allow us to isolate the
effects of the constant, $W_0$, appearing as follows:
\begin{align}
  W= h\mbox{Tr}\left(\phi \Phi \tilde{\phi}\right) - h \mu^2
  \mbox{Tr}(\Phi) + h W_0
\end{align}

The presence of this constant slightly changes the global SUSY
minimum, introducing a modest amount of SUSY breaking.  One can tune
the constant such that the superpotential vanishes with the F-terms,
but it is not possible to achieve this if we wish to have $V=0$ at the
metastable point.

The constant creates an AdS minimum with a negative expectation value
equal in magnitude to the global ISS theory's, namely $V_{\mbox{ADS}}
\simeq -h^2 N_c \mu^4$.  However, the difference between $V$ in the AdS minimum
and in the metastable minimum is essentially the same as the the
difference between $V$ in the SUSY minimum and in the metastable in
the global case.  The height of the barrier is also essentially the
same in both cases.

Finally our numerical studies demonstrate that if $W_0\sim \mu^2$ then
$\Phi_0$ gets expectation values of order 1, but the expectation value
shrinks as $W_0 \rightarrow 0$, going to zero in that limit.

\section{One loop potential}\label{sec:One Loop}

Since the interplay between the supergravity and one-loop effects is
so important to our results it is worth discussing the details of the
one-loop calculation, highlighting the approximations we have made.
First of all we note that the mass matrices, $\mathcal{M}$, in the
well-known formula:
\begin{align}
  V_{\mbox{one-loop}} = V_{\mbox{quad.}}+ V_{\mbox{log}} =
  \frac{1}{32\pi^2}\mbox{STr}\mathcal{M}^2 \Lambda^2 + \frac{1}{64
    \pi^2}\mbox{STr}\mathcal{M}^4\log\frac{\mathcal{M}^2}{\Lambda^2}
  \label{eq:V_one-loop}
\end{align}
are given by the supergravity corrected masses \cite{Chung:2003fi,
  Cremmer:1982en}.  This modifies the mass squared matrices at the
level of $\mu^4$ (i.e.~a rather small shift, but calculable and
necessary for the computation of $V_{\mbox{quad.}}$). The term
quadratic in $\mathcal{M}$ is generically $\sim m_{3/2}^2 \Lambda^2$
whereas the contribution from $V_{\mbox{tree}}$ is $-3 m_{3/2}^2$,
hence the $\mathcal{M}^2$ term can be disregarded if $\Lambda \ll
M_{P}$, but not otherwise.  Unfortunately Eq.~(\ref{eq:V_one-loop}) is
modified when $V\neq 0$ (for discussions of this point, see
\cite{deAlwis:2006nm} and \cite{Gaillard:1993es}).  Even though we are
expanding about $V=0$, there will be corrections to this expression
since $V=0$ is only true at that point and in the Goldstone
directions.  We expect there to be both logarithmic and quadratic
corrections stemming from this. The quadratic terms we can ignore if
$\Lambda \ll 1$, but the logarithmic terms we have to consider more
carefully.  We note that all the operators in \cite{Gaillard:1993es}
that contain $V$ are dimension 8 and so the largest possible linear
contribution would be $\mathcal{O}(W_0^2 \mu X)\sim \mu^5X$ where $X$
is a generic field.  This should be compared to the largest
contribution from gravity at tree-level, $X F W_0\sim \mu^4 X$, and so
we expect these effects to be negligible.

As noted in \cite{Intriligator:2006dd} and \cite{Grisaru:1996ve} one
can capture some information about the effective potential purely by
integrating out fields and calculating the correction to the K\"ahler
potential.  However, as described in the appendix of
\cite{Intriligator:2006dd}, this is an approximation only valid to 2nd
order in $F$, we also note that it is harder to work with numerically.
Hence we opt to calculate the full one-loop potential.  It is
nonetheless interesting to compare these two approaches and we see
that the (somewhat arbitrary) corrections to the K\"ahler potential
introduced in \cite{Haba:2008vj}, created an explicit cut-off
dependence into the effective potential for $\Phi_0$ and hence
$\<\Phi_0\>\propto \Lambda^2$.  This dependence is not present in
global theory and we found that only a very mild dependence was
introduced by including supergravity corrections to the logarithmic
effective potential, as we demonstrate in section~\ref{sec:Numerics}.
Regrettably, this does not provide a rigorous test of the two
approaches, due to the K\"ahler corrections being more postulated than
derived in \cite{Haba:2008vj}.

It is then not entirely surprising that our results differ markedly
from those of \cite{Haba:2008vj}.  This manifests itself primarily in
our predictions for the expectation value of $\Phi_0$ which we find to
be significantly smaller than $\mu$, irrespective of the value of the
cut-off.  This means that our model does not appear to be a good
candidate for gauge mediation, since $\<\Phi_0\>^2<F_{\phi_0}$.
However, the gravitational corrections to the quarks F-terms open the
possibility that they could couple to a mediation sector and generate
soft terms.  We will return to this in section \ref{sec:SoftMasses}.

\subsection{Analytic properties of STr$\mathcal{M}^2$}

For a canonical K\"ahler potential, the quadratic one-loop potential is given by
\begin{align}
V_{\mbox{quad.}}=\frac{\mbox{STr}\mathcal{M}^2\Lambda^2}{32\pi^2} =   \frac{\Lambda^2 e^K}{16\pi^2} (N_f^2+2 N_f N_c-1)\left(\sum_i \left(W_i+X_iW \right)^2 \right) -  \frac{\Lambda^2 e^K}{8\pi^2} (N_f^2+2 N_f N_c) W^2.
\end{align}
Where fields are taken to be real and $X_i$ runs over all fields.  Re-writing this in terms of the tree-level potential gives:
\begin{align}
 V_{\mbox{quad.}} = \frac{\Lambda^2}{16\pi^2} (N_f^2+2 N_f N_c-1) V_{\mbox{tree}} + \frac{\Lambda^2e^K}{16\pi^2} ( ( N_f^2+2 N_f N_c) - 3) W^2 \label{eq:msquared}
\end{align}
and hence
\begin{align}
  V_{\mbox{tree}}+ V_{\mbox{one-loop}} = \left((N_f^2+2 N_f N_c - 1) \frac{\Lambda^2}{16\pi^2} + 1 \right) V_{\mbox{tree}} + \frac{\Lambda^2}{16\pi^2} ( N_f^2+2 N_f N_c - 3) e^K W^2 + V_{\mbox{log}} \label{Eq:One-loop breakdown}
\end{align}

The addition of $V_{\mbox{quad.}}$ to the potential reinforces the
tree-level solution, up to additional, gravitationally suppressed
contributions from the final term in Eq.~(\ref{eq:msquared}).

On dimensional grounds, the SUSY parts of the Sugra F-terms will
provide the dominant contributions to $V_{\mbox{quad.}}$, except for
the moduli fields, which have flat F-terms at the SUSY minimum.  This
means we expect the minima of $V_{\mbox{quad.}}$ and
$V_{\mbox{tree}}$, for the non-moduli fields, to coincide at leading
order in $\mu$ (assuming the moduli are taken to be $\sim \mu^2$).
However, the gravitational corrections given by $W^2$ and by 
$K_i W$ come in at the same order of magnitude and hence we do not
expect that the same $\Phi_0$ will minimise both $V_{\mbox{quad.}}$
and $V_{\mbox{tree}}$.  It is nonetheless clear from
Eq.~(\ref{eq:msquared}) that, if $(N_f^2+2 N_f
N_c)\frac{\Lambda^2}{16\pi^2} \sim 1$, the quadratically divergent
corrections will be of equal importance to the tree-level.

We may also compute the value of $W_0$ required to cancel the
cosmological constant:
\begin{align}
 W_0=\left(\frac{e^{-K} V_{\mbox{log}} + \left(N_f'\Lambda'^2 +1 \right)h^2 N_c \mu^2}{3\left(N_f'\Lambda'^2+1\right)-(N_f'-2)\Lambda'^2} \right)^{1/2}
\end{align}
Where $\Lambda'=\frac{\Lambda}{4\pi}$ and $N_f'=N_f^2+2 N_f N_c-1$.

The minima of the logarithmic potential and tree-level potential need
not coincide, which is fortunate since the tree-level potential is
minimised by Planck scale moduli vevs. Therefore the addition of the
logarithmic potential to the tree-level will shift the minimum away
from the tree-level, with the size of the shift being determined by
the strength of the coupling constants and the relative importance of
gravity.  Inclusion of the quadratic, one-loop
potential will shift the minimum closer to the tree-level result for
all fields as discussed in appendix~\ref{sec:appB}.

\subsection{Remarks on methodology}

Since computation of $V_{\mbox{log}}$ requires diagonalisation of
$\mathcal{M}^2$ it is significantly more involved than the calculation of
the quadratic piece.  It is possible to make some analytic progress by
using the one-loop moduli masses derived in \cite{Intriligator:2006dd}
and tree-level gravity corrections.  However, more involved analytic
calculations, such as computing the logarithmic piece of
Eq.~(\ref{eq:V_one-loop}) (with or without gravitational corrections
to the masses) or computing corrections to the K\"ahler potential, as
detailed in \cite{Grisaru:1996ve} are extremely challenging since they
both rely upon diagonalization of large mass matrices.  We can make
some approximate analytical statements by observing that the direct
gravitational corrections to the logarithmic effective potential for
$\Phi_0$ given in \cite{Intriligator:2006dd} are small and
overwhealmed by the non-gravitational terms.  Then using the global
effective potential, derived at the global SUSY minimum, will only
introduce small errors if it does not vary too rapidly across field
space and the combination of this and the Sugra tree-level potential
gives vevs close to the global vevs.  To see if this approximation was
valid we opted to calculate the effective potential numerically, using
the approach described in the following section. One can make progress
by observing that, at the minimum, it is possible to compute the
series expansion the one-loop potential up to second order in the
fields.  This leaves enough information to confirm that this is point
is both a stationary point and a minimum.  Eq.~(\ref{eq:V_one-loop})
can then be computed for pairs of fields at a time, with all others
frozen.  With the problem broken into several managable parts it is
possible to attack it numerically as described in the following
section.

However computing the effective K\"ahler potential is a more difficult
task, since one needs to know the series expansion of the matrix of
second derivatives of the inverse K\"ahler potential up to second
order in the fields.  This is in addition to knowing the K\"ahler
potential up to second order. To calculate the second derivatives of
the inverse K\"ahler potential to second order we would need to
calculate the K\"ahler potential to fourth order, doubling the number
of fields we have to consider simultaneously.  This alone shows that this
approach should be significantly more time consuming.

As a technical point we note that the Coleman-Weinberg contribution to
$\Phi_0$'s mass derived in ISS is not valid away from the global ISS
minimum.  If we only include the one-loop masses derived in
\cite{Intriligator:2006dd} and
allow 
all fields to vary, then the numerical solution has negative
eigenvalues in the Hessian.  This apparent instability can be shown to
be an artifact caused by the mismatch between the global minimum (the
point at which ISS's masses are correct) and the true minimum (in
which they are not).  While these effects are small we can
re-calculate the one-loop contribution at the minimum of the
tree-level plus one-loop potentials, we can trust our results.
Unfortunately, since $\<\Phi\>\neq 0$ and it is no longer a simple
matter to determine the Goldstone directions, the calculation of the
masses must be done for all fields and becomes significantly more
involved \footnote{This is in contrast to global ISS, or indeed ISS
  with no additional physics, because in our case $\<\Phi\>\neq 0$.
  When $\<\Phi\>=0$ the Goldstone bosons are solely linear
  combinations of the magnetic quarks, but when $\<\Phi\>\neq 0$ it
  contributes to the breaking of $SU(N_f)$ and hence the Goldstone
  bosons.  This means that any contributions to the Goldstone bosons'
  potential that violate the original $SU(N_f)$ symmetry give
  (spurious) masses to the Goldstones.  Hence one should compute the
  full potential, at least to 2nd order in the fields, to obtain
  reliable results.}. As a result of this, we discovered non-trivial
contributions to not only $\Phi_0$'s potential, but $\phi_0$ and
$\Phi_1$'s.

Our approach was to start at a point close to the global ISS minimum,
compute the one-loop potential to second order in the fields at this
point and then minimise the tree-level + one-loop potential
\footnote{This avoids the need to diagonalise the mass matrix
  analytically.  Which would be challenging because it is both large
  and has numerous independent variables.  Since we only need the
  local properties of the potential we only need calculate the its
  numerical values after small variations in the fields. If one wanted
  to know the values of the potential for all values of the fields, it
  would be necessary to diagonalise analytically.}.  This new point
was then used to re-compute the one-loop potential, allowing us to
minimise yet again.  This process was repeated until the solution
converged and the eigenvalues of the matrix of second derivatives of
the full potential were consistent with the expected number of
Goldstones (which would not be the case if the effective potential
were computed away from the minimum).

Regrettably, this is not sufficient to make the computation tractable
for large numbers of flavours, as the computation time grows rapidly
with the number of flavours and $N_f=4$, $N_c=3$ is already takes
prohibitively long.  However, for $N_f=3$, $N_c=2$ the one-loop
potential can be computed in a reasonable amount of time.

\section{Non-perturbative contributions}\label{sec:NonPert}

To estimate the value of the cut-off for which the non-perturbative
piece (\ref{nnpert}) dominates, and destabilises the potential, we calculate the second
derivative of the non-perturbative correction to the potential,
evaluated at the minimum of the tree-level + one-loop effective
potential. 
If $\frac{\partial^2
  V_{\mbox{non-pert.}}}{\partial \Phi^2} >\frac{\partial^2\left(
  V_{\mbox{tree}}+V_{\mbox{one-loop}}\right)}{\partial \Phi^2}$, then the non-perturbative piece
will likely dominate and the fields will roll to the supersymmetric
minimum.  This allows us to put rough lower bounds on the cut-off,
such that the meta-stable solution is stable.  This is in
contrast to the global case in which the non-perturbative effects
vanish in the tree-level + one-loop minimum.  Considering, for
simplicity's sake, $N_f=N_c+1$:
\begin{align}
  \frac{\partial^2V_{\mbox{non-pert.}}}{\partial \Phi_1 \partial (\Phi_0)_i} \sim \sum_j^{N_f-1} 2 h^{N_f/N} \Lambda_m^{-(N_f-3)}  \left( \prod_{k\neq j,i}^{N_f-1} (\Phi_0)_i \right)(-h \mu^2)
\end{align}
where all higher powers of $\Phi$ have been discarded.

Close to the global SUSY minimum the dominant contribution to the effective potential
is the globally supersymmetric effective potential derived in
\cite{Intriligator:2006dd}.  This means that

\begin{align}
  \frac{\partial^2 V_{\mbox{log}}}{\partial \Phi_{ii}^2}\sim \frac{\log(4)-1}{4\pi^2}h^4\mu^2
\end{align}
and hence if $\sum_j 2 h^{\frac{N_f+N}{N}} \Lambda_m^{-(N_f-3)} \left(
  \prod_{k\neq j,i}^{N_f-1} (\Phi_0)_i \right)>
\frac{\log(4)-1}{4\pi^2}h^4$ it is clear that the non-perturbative
piece will dominate and the fields will evolve into the SUSY minimum.
However, we can see that, even if the cut-off has the same order of
magnitude as $\mu$, the non-perturbative potential alone will not have
a significant effect.  It is suppressed by the small expectation
values of $\Phi_0$, given by the perturbative potential.  If $\<\Phi\>
\gg \mu$ the non-perturbative potential can dominate, but this is far
from the case here.

We also note that, while there are non-perturbative contributions to
the one-loop effective potential, these effects are generically small.
They will only need close consideration when 
$\<\Phi\> \gg \mu$ and even in this case, the
non-perturbative contributions to the tree-level will be more
important, except at singular points where the perturbative expansion
breaks down.

Finally, the complete calculation with a fully realistic number of
flavours is numerically intractable, and we are only able to compute
everything for the case where $N_f=3$ and $N_c=2$.  While this case
does not correspond to a dualized theory, it does capture the
important low energy phenomena: the rank condition still holds and
SUSY is still broken spontaneously.  However, the non-perturbative
contributions are qualitatively different.  Firstly the coefficient is
automatically 1, irrespective of the size of the cut-off, and
the determinant piece is larger by roughly $h^2 \mu^{-2}$, since it
contains one fewer power of $\Phi_{ij}$.  This means that the
non-perturbative piece can come to dominate, even though we know that
it would be negligible in the $N_f=4$, $N_c=3$ case.  As a result, we
were forced to introduce a constant, $\Lambda_{\mbox{non-pert.}}$,
multiplying the non-perturbative piece and find the largest stable
value, before we could be certain that the non-perturbative effects
were under control.
More specifically
\begin{align} 
W_{non-pert} \rightarrow   W_{non-pert} = \Lambda_{\mbox{non-pert.}} N h^{N_f/N}\bigl(\Lambda_m^{-(N_f-3N)}\det(\Phi)\bigr)^{1/N}
\end{align}
in what follows.

\section{Supersymmetry Breakdown}
\subsection{Numerical Results}\label{sec:Numerics}

  \begin{table}
  \centering
   \begin{math}
     \begin{array}[b]{|c|ccccccc|}
       \hline
       N_f  & 3 & 3 & 3 & 3 & 3 & 3 & 3 \\
       N  & 1 & 1 & 1 & 1 & 1 & 1 & 1 \\
       \mu   & ~~1.41~10^{-7} & ~~1.41~10^{-7} & ~~1.41~10^{-7} & ~~1.41~10^{-7}  & ~~1.41~10^{-7} & ~~1.41~10^{-7} & ~~1.41~10^{-7}  \\
       \Lambda_{\mbox{non-pert.}}& 0 & ~~2.00~10^{-6} & ~~2.00~10^{-4}  & ~~2.00~10^{-2} & ~~4.00~10^{-2} & ~~4.40~10^{-2} & ~~4.60~10^{-2}  \\
       \hline
       m_{3/2} & ~~1.52~10^{-14}  & ~~1.52~10^{-14} & ~~1.52~10^{-14} & ~~1.52~10^{-14} & ~~1.52~10^{-14} & ~~1.52~10^{-14} & ~~1.52~10^{-14} \\
       \<\phi_0\> & ~~1.41~10^{-7~}   & ~~1.41~10^{-7~} & ~~1.41~10^{-7~} & ~~1.41~10^{-7~} & ~~1.41~10^{-7~} & ~~1.41~10^{-7~} & ~~1.41~10^{-7~} \\
       \<\Phi_0\> & - 6.33~10^{-12} & -6.33~10^{-12} & -6.33~10^{-12} & -7.95~10^{-12} & -1.57~10^{-11} & -2.10~10^{-11} & -2.56~10^{-11}  \\
       \<\Phi_1\> & -1.21~10^{-13} & -1.21~10^{-13} & -1.23~10^{-13} & -3.71~10^{-13} & -1.16~10^{-12} & -1.66~10^{-12} & -2.08~10^{-12}  \\
       \hline
       F_{\phi_0}/\mu^2 & -7.43~10^{-7~}  & -7.43~10^{-7~} & -7.56~10^{-7~} & -2.50~10^{-6~} & -8.03~10^{-6~} & -1.15~10^{-5~} & -1.45~10^{-5~}  \\
       F_{\Phi_1}/\mu^2 & -1.07~10^{-2~} & -1.07~10^{-2~} & -1.07~10^{-2~} & -1.07~10^{-2~} & -1.07~10^{-2~} & -1.07~10^{-2~} & -1.07~10^{-2~} \\
        \hline
      \end{array}
   \end{math}
   \caption{Solutions for V=0, with the non-perturbative piece included.  The dependence on the non-perturbative contribution is shown to be very small, but as $\Lambda_{\mbox{non-pert.}}$ exceeds $2~10^{-2}$ it rapidly comes to dominate.  The value of $m_{3/2}$ is identified with $W_0$, up to small corrections, supressed by additional powers of $\mu$.}
   \label{table:Non-pert}
 \end{table}

 We now present the numerical analysis of our model.  In the
 following we confirm that the metastable minimum exists in the
 presence of gravity, show where the numerical results diverge from
 the analytical approximations and explain why these deviations are
 larger than dimensional analysis suggests.  Where possible we compare
 our results with those presented previously, demonstrating that they
 can be recovered if the same assumptions about the one-loop
 potential are made (the assumptions used to derive the analytical
 approximations), but that relaxing these assumptions introduces the shift
 just discussed.

 Our main observation in this section is that one must be careful
 about estimating the error introduced by neglecting gravitational
 corrections to the one-loop effective potential.  As far as the
 logarithmic one-loop potential is concerned, it is necessary to
 compute it in full, though the gravitational corrections can safely
 be ignored \footnote{We included them in our numerical analysis for
   completeness, but the corrections proved to be small.}.  The
 quadratic one-loop potential must be included.  While the quadratic
 one-loop potential is zero in the absence of gravity, the
 gravitational corrections generate a potential similar to the
 gravitational corrections to the tree-level, but controlled by an
 overall factor of $\frac{\Lambda^2}{32\pi^2}$ - as we show in appendix
 \ref{sec:appB}.  These potentials, combined with the tree-level, must
 be computed in order to obtain a reliable leading order estimate in
 general, with the quadratic potential being of particular interest if
 the number of fields is large and the cut-off close to the Planck
 scale.

 In table~\ref{table:Non-pert} the row $\Lambda_{\mbox{non-pert.}}$ is
 the coefficient of the non-perturbative piece, introduced to
 compensate for the missing powers of $\Phi_{ij}$ that would be
 present in the determinant for a realistic number of flavours.  Note,
 the F-term for $\Phi_0$ is not included since it is only shifted by
 corrections of order $\mu^3$, so $F_{\Phi_0}=h
 \mu^2+\mathcal{O}(\mu^3)$.  For physically reasonable values of $\mu$
 this effect is negligible.  Also, $\Lambda=10^{-2}$, (i.e. the
 string/GUT scale) throughout tables~\ref{table:Non-pert} and
 \ref{table:Mu dep}.  The reason is, this value of $\Lambda$ is
 sufficiently small and ensures that the quadratically divergent loop
 correction to the potential only creates a small shift from the
 tree-level result, leaving us with the logarithmic piece whose
 sensitivity to the value of the cut-off is very
 weak.

 Since the superpotential can be written $W=W_{ISS}+ h W_0$, and
 $\<W_{ISS}\>=0$ in the global limit, we find that $ m_{3/2} = e^K h
 W_0 + \mathcal{O}(\mu^3) \simeq h W_0$.  Hence we only include
 $m_{3/2}$ in the tables.  Also, in all the tables $h$ is taken to be 1.
 Our main result, concerning the soft mass spectrum, is independent of
 the precise value of $h$, though we retain full theoretical control
 in a surprisingly small range.  If $h\ll 10^{-1}$ the $\Phi_0$
 modulus runs off to the Planck scale as the gravitational effects
 overwhelm the one-loop contributions (since $\mu$ must go like $\mu
 \rightarrow \frac{\mu}{h}$, as $h$ varies from 1, and hence
 gravitational effects increase in relevance as $h$ shrinks).
 However, if $h\gg 1$ the theory becomes strongly coupled.

 Our numerical studies show that the non-perturbative piece is under
 control, that there is only a very mild, logarithmic cut-off
 dependence, when supergravity corrections are accounted for, and that
 the main features of the model are independent of $\mu$, assuming
 $\mu \ll M_P$.  These results can be seen in, respectively,
 tables~\ref{table:Non-pert}, \ref{table:lam dep} and \ref{table:Mu dep}.

 From the data in table~\ref{table:lam dep} we can see that for
 $\mu=\sqrt{2}~10^{-7}$, $\<\Phi_0\>= -4.295~10^{-12} + 6.886~10^{-14}
 \mbox{Log}_{10} (\Lambda)$.  Table~\ref{table:Mu dep} shows that
 there are no significant changes as one varies $\mu$. In
 table~\ref{table:Non-pert} $\Lambda_{\mbox{non-pert.}}$ is varied and
 the non-perturbative term becomes relevant for
 $\Lambda_{\mbox{non-pert.}}>4~10^{-2}$.  Since we expect the
 non-perturbative term to be suppressed by an extra power of $\Phi$
 when $N_f=4$, corresponding to $\Lambda_{\mbox{non-pert.}}\sim \Phi
 \sim \mu^2 = 2~10^{-14}$ it is clearly well under control.

\subsection{Analytic Approximation}
In addition we can compare our numerical results to analytical
approximations.  Because of the expected smallness of the fields it is
sufficient to take the leading order of $\mu$ when searching for the
minimum.  The gravitational corrections to the logarithmic one loop
potential appear with higher powers of $\mu$ than the SUSY effective
potential and can be neglected in the analytical approximation, while
the quadratic corrections are both relevant and readily calculable.
The solution to $\frac{\partial V}{\partial \Phi_0}=0$, derived in
appendix~\ref{sec:appB}, under the assumption\footnote{That the one-loop
  potential is given entirely by the globally supersymmetric one-loop
  potential derived in \cite{Intriligator:2006dd}.} that $V_{\mbox{log}}$ is
described by Eq.~\eqref{eq:v_log_ISS}, is given by:

\begin{table}
  \centering
   \begin{math}
     \begin{array}[b]{|c|ccccccc|}
       \hline
       N_f & 2 & 2 & 2 & 2 & 2 & 2 & 2\\
       N   & 1 & 1 & 1 & 1 & 1 & 1 & 1\\
       \mu  & ~~1.41~10^{-7~} & ~~1.41~10^{-7~}  & ~~1.41~10^{-7~} & ~~1.41~10^{-7~} & ~~1.41~10^{-7~} & ~~1.41~10^{-7~}  & ~~1.41~10^{-7~}  \\
       \Lambda  & 1 & 10^{-1} & 10^{-2} & 10^{-3} & 10^{-4} & 10^{-5} & 2~10^{-7}\\
       \hline
       m_{3/2} & ~~1.04~10^{-14} & ~~1.06~10^{-14} & ~~1.08~10^{-14} & ~~1.09~10^{-14} & ~~1.11~10^{-14} & ~~1.13~10^{-14} & ~~1.16~10^{-14}  \\
       \<\phi_0\> &  ~~1.41~10^{-7~}  & ~~1.41~10^{-7~}  & ~~1.41~10^{-7~}  & ~~1.41~10^{-7~} & ~~1.41~10^{-7~} & ~~1.41~10^{-7~} & ~~1.41~10^{-7~}  \\
       \<\Phi_0\> & -4.29~10^{-12}  & -4.36~10^{-12}  &  -4.43~10^{-12} & -4.50~10^{-12}  & -4.57~10^{-12} & -4.64~10^{-12} &  -4.75~10^{-12} \\
       \<\Phi_1\> & -4.78~10^{-14}  & -4.44~10^{-14} & -4.16~10^{-14} & -3.91~10^{-14}  & -3.70~10^{-14}  & -3.51~10^{-14}  &  -3.25~10^{-14}\\
       \hline
       F_{\phi_0}/\mu^2 & -2.64~10^{-7~} & -2.39~10^{-7~}  & -2.17~10^{-7~} & -1.99~10^{-7~}  & -1.83~10^{-7~}  & -1.68~10^{-7~}  & -1.47~10^{-7~}   \\
       F_{\Phi_1}/\mu^2& -5.66~10^{-3~}  & -5.46~10^{-3~}  & -5.27~10^{-3~} &  -5.09~10^{-3~}  & -4.93~10^{-3~}  & -4.77~10^{-3~}  & -4.53~10^{-3~} \\
       \hline
      \end{array}
  \end{math}
  \caption{Solutions for V=0 where the non-perturbative and one-loop quadratic pieces have been neglected.  These data show the logarithmic dependence of $\Phi_0$ on the cut-off.} 
   \label{table:lam dep}
\end{table}

\begin{align}
 \<\Phi_0 \>= \frac{ 16\pi^2 W_0 (1+\frac{\Lambda'^2}{2}(N_f'+2))}{h^2(\mbox{log}(4)-1)} +  \mathcal{O} (\mu^4) \label{Eq:AnotherPhi0}
\end{align}

Comparing this with the numerical results, obtained using the full
one-loop potential, shows that they disagree at the percent level.
For example, it gives $\<\Phi_0\>=-4.25~10^{-12}$, with $N_f=2$,
$\mu=\sqrt{2}~10^{-7}$, $\Lambda=1$ and $V_{\mbox{quad.}}$ neglected
(i.e. $\Lambda'=0$), to be compared with table~\ref{table:lam dep}.
Since the difference is far greater than the error we expected to be
introduced by ignoring gravitational corrections to the one-loop
potential, we conclude that the assumption that the one-loop potential
is given by Eq.~\eqref{eq:v_log_ISS} is not justified in the
presence of gravity.  This deviation
is a consequence of the changes in the effective potential for
$\Phi_0$ caused by moving from the global SUSY minimum to the Sugra
minimum, in $\Phi_1$-$\phi_0$ space. If we artificially shift all
fields except $\Phi_0$ to their global SUSY vevs then the Sugra
corrected logarithmic effective potential for $\Phi_0$ tends to the
SUSY effective potential.  The gravitational corrections to
$V_{\mbox{log}}$ at the global SUSY minimum are negligible, as
expected.  It is also interesting to see that at this point in
$\Phi_1$-$\phi_0$ space the one-loop correction to the potential
contains a term linear in $\Phi_0$ and hence is minimised by
$\<\Phi_0\>\neq 0$.

We also notice similar behaviour when $V_{\mbox{quad.}}$ is included.
For $N_f=2$, $\mu=\sqrt{2}~10^{-7}$, $\Lambda=1$ Eq.~(\ref{Eq:AnotherPhi0})
predicts $\<\Phi_0\>=-4.37~10^{-12}$, but the minimum appears at
$\<\Phi_0\>=-4.46~10^{-12}$

As noted in appendix~\ref{sec:appA}, when $W_0=0$, the expectation
values of the quarks are determined by the global minimisation
conditions modified by the expectation value of the tree-level
potential.  Cancelling the cosmological constant at tree-level, via
$W_0$, recovers the global result, up to small corrections induced by
$W_0$. When the loop corrections are removed, $\Phi\rightarrow 0$ and $N_c=1$ we find

 \begin{table}
  \centering
   \begin{math}
     \begin{array}[b]{|c|cccc|}
       \hline
       N_f  & 2 & 2 & 2 & 2 \\
       N   & 1 & 1 & 1 & 1\\
       \mu  & 10^{-7} & 10^{-6} & 10^{-5} & 10^{-4} \\
       \hline
       m_{3/2}  & ~~5.36~10^{-15} & ~~5.45~10^{-13} & ~~5.54~10^{-11} & ~~5.63~10^{-9~} \\
       \<\phi_0\> & ~~9.97~10^{-8~} & ~~9.97~10^{-7~} & ~~9.98~10^{-6~} & ~~9.98~10^{-5} \\
       \<\Phi_0\> & -2.21~10^{-12} & -2.25~10^{-10} & -2.28~10^{-8~} & -2.32~10^{-6~} \\
       \<\Phi_1\> & -2.10~10^{-14} & -1.97~10^{-12}& -1.86~10^{-10} & -1.77~10^{-8~} \\
       \hline
       F_{\phi_0}/\mu^2 & -1.56~10^{-7~} & -1.42~10^{-6~} & -1.31~10^{-5~} & -1.20~10^{-4~} \\
       F_{\Phi_1}/\mu^2 & -5.30~10^{-3~} & -5.12~10^{-3~} & -4.95~10^{-3~} & -4.79~10^{-3~} \\
       \hline
      \end{array}
  \end{math}
  \caption{Solutions for V=0 where the non-perturbative and one-loop quadratic pieces have been neglected. Here, the $\mu$ dependence is shown.} 
   \label{table:Mu dep}
\end{table}

\begin{align}
  \<\phi_0\>^2 & =\frac12\left( -1 +2 \mu^2 -2 W_0^2+ \sqrt{1- 4 N_c\mu^4+12 W_0^2 - 8 \mu^2 W_0^2 +4 W_0^4}\right) \\ 
  & = \mu^2 - N_c \mu^4 + 2  W_0^2 - 2 \mu^2 W_0^2 +\mathcal{O}(\mu^8) \label{eq:phi_square}
\end{align}
hence, to good approximation, $\<\phi_0\>^2=\mu^2$ for
$\mu=\sqrt{2}~10^{-7}$, in agreement with the leading order
calculation in appendix~\ref{sec:appB}.  At $\mu^4$ order this result
depends on the cancellation of the cosmological constant, requiring
the equality of the $F$ term contribution and $-3 W^2$.  Hence, the
one loop potential would shift $\<\phi_0\>^2$ by $\lesssim \mu^4$,
even if it were independent of $\phi_0$ (as has tacitly been assumed
in \cite{Serone:2007sv}, \cite{Abe:2006xp}, \cite{Dudas:2006gr} and
\cite{Xu:2007az}) and merely contributed to the cosmological constant.
Moreover, the one-loop potential proves to have a non-trivial
dependence on $\phi_0$ and, for $h=1$, introduces a correction at the
level of $\frac{\mu}{1000}$.  It should be stressed that this effect
remains when the one-loop potential is purely supersymmetric, as long
as gravity is switched on in the tree level potential.  For
$\Lambda=1$, $N_f=2$ and $N_c=1$ we find $\<\phi_0\>=1.41042~10^{-7}$
if $ V_{\mbox{quad.}}$ is present and $\<\phi_0\>=1.41020~10^{-7}$ if
it is not. In contrast, these two results would be indistinguishable
at this level of precision had only the logarithmic corrections to
$\Phi_0$'s potential been accounted for.

Similarly to $\Phi_0$, $\Phi_1$'s logarithimic one-loop potential depends on the
moduli expectation values and, unlike the quarks, does not tend to the
tree-level result as $h\rightarrow 0$ (and $V_{\mbox{quad.}}$ is
neglected).  For $N_f=2$, $\mu=10^{-7}$, $\Lambda=1$ and with
$V_{\mbox{quad.}}$ neglected we obtain $\<\Phi_1\>=-4.78~10^{-14}$
which differs by a factor of 5 compared to the leading order
tree-level result of $\<\Phi_1\>=-W_0=-1.04~10^{-14}$ derived in
appendix~\ref{sec:appB}.  The results with $V_{\mbox{quad.}}$ included are
closer as the tree-level + quadratic potential also gives
$\<\Phi_1\>=-W_0=-1.04~10^{-14}$, and the full potential gives
$\<\Phi_1\>=-4.65~10^{-14}$.

\section{Soft Masses}\label{sec:SoftMasses}
It is well known that supergravity theories automatically include the
gravitational mediation mechanism and soft-terms will be generated.
The typical scale for these soft masses is $m_{3/2}$ with deviations
being generated by non-trivial K\"ahler potentials.  Also, gaugino
masses can be zero at tree-level, if the gauge kinetic function
preserves supersymmetry.  Since we know the gravitational
contributions must be present we now analyse the relative importance
of gauge and anomaly mediation and sketch the features of the
spectrum.

Since the addition of a constant to the superpotential implies that
$\<\Phi\>\neq 0$, in supergravity, we investigated the possibility that
the R-symmetry violating, non-perturbative piece could give masses
to the gauginos.  The determinant piece (\ref{nnpert}) can have a non-trivial
contribution to the ISS fields' masses and, if we employed a direct
mediation mechanism analogous to, for example, \cite{Abel:2007jx}, the
R-symmetry breaking could in principle be transfered to the MSSM sector.
However, as we saw in table~\ref{table:Non-pert}, the square of the scalar vev of $\Phi$ is less
than the F-term.  Hence, $\Phi$ cannot be the '$X$' field \footnote{We
  have in mind an operator $W\ni X f \overline{f}$ where $f$ and
  $\overline{f}$ are messenger fields charged under the visible sector
  gauge groups.} that couples to the messengers since, if $\<\Phi\>=M +
F \theta^2$ then $M^2 < F$, and, if $\<\Phi\>$ is the sole contributor
to the messenger masses, then they will be tachyonic (as shown in
\cite{Giudice:1998bp}).

However, the gravitational effects induce new F-terms that are not
present in global SUSY.  Being generated gravitationally, they are
always smaller than the F-terms of $\Phi_0$, assuming the
gravitational effects are under control \footnote{Since the
  gravitational effects come in with an additional power of $\mu$
  compared with the global SUSY terms and the new F-terms are $\sim h
  \mu^3$.}.  Since the magnetic squarks' vevs are $\sim \mu$ and their
F-terms $\sim h \mu^3$ they automatically satisfy the $M^2>F$
condition, if $h\leq 1$.  An example messenger sector would have the
symmetry group $SU(N) \times SU(N_f) \times SU(5) \times U(1)_R$ and
the fields would transform as follows: $\phi_0:(N,N_f,1,0)$,
$f:(\bar{N},1,5,1)$, $f':(1,\bar{N_f},\bar{5},1)$.

The crucial observation is that two SUSY breaking scales are generated
automatically, if $V=0$ is required.  Hence the standard argument for
the dominance of gauge mediation, namely that $\frac{F}{M}\gg
\frac{F}{M_P}$, does not necessarily apply.  Instead we have
$F_{\Phi_0}\gg F_{\phi_0}$ and $\frac{F_{\phi_0}}{\<\phi_0\>} \sim
\frac{F_{\Phi_0}}{M_P}$.  While it is true that
$\frac{F_{\Phi_0}}{\<\Phi_0\>}\gg \frac{F_{\Phi_0}}{M_P}$, $\Phi_0$
cannot be allowed to couple to the messengers because
$F_{\Phi_0}>\<\Phi_0\>^2$.

The naive expression for the gaugino masses is given by:

\begin{align}
  m_{\lambda}\sim\frac{\alpha}{4\pi} \frac{F_{\phi_0}}{\<\phi_0\>} \sim
  \frac{\alpha}{4\pi} \frac{h \<\phi_0\>W_0}{\<\phi_0\>} \sim \frac{\alpha}{4\pi} h \mu^2
  \label{eq:gaugino_mass}
\end{align}
and the soft scalar mass squareds are approximately:
\begin{align}
  m^2=m_{\lambda}^2.
\end{align}

In addition we can estimate the soft mass contributions from anomaly mediation:

\begin{align}
  M_a=F_{\mbox{Anom.}}\beta_{g_a}/g_a \sim \frac{\alpha}{4\pi} F_{\mbox{Anom.}}
\end{align}
\begin{align}
  (m^2)^i_j=F_{\mbox{Anom.}}^2 \frac{\partial \gamma^i_j}{\partial t} \sim \frac{\alpha^2}{16\pi^2}F_{\mbox{Anom.}}^2
\end{align}

Since $F_{\mbox{Anom.}}$ can at most be $F_{\Phi_0}$ (without
postulating another source of SUSY breaking) it follows that these
contributions are of the same order of magnitude as those given by
gauge mediation.

This allows us to estimate the size of the gravitino mass, based on
the requirement that the gaugino masses be in the vicinity of a TeV and
that the cosmological constant be tuned to zero.  These requirements
imply that $\frac{\alpha}{4\pi}h \mu^2\sim 1~ \mbox{TeV} \sim
10^{-16}$ and thus $h\mu^2\sim \frac{4\pi}{\alpha} 10^{-16}$.  Taking
$\alpha=\frac{1}{26}$ \cite{Kazakov:2000ra} we find $h \mu^2 \sim
3~10^{-14}$, up to order one factors, and we see that $m_{3/2}\simeq
e^K h W_0 \simeq \left(\frac{h^2 N_c \mu^4}{3} \right)^{1/2}$ and hence the
gravitational contribution to the soft scalar masses is automatically
two orders of magnitude larger than the other contributions.

As mentioned earlier, the gravitational contribution to the gaugino masses is not
necessarily order $m_{3/2}$.  For example in string theories one can
find that the gauge kinetic functions, whose expectation values
specify the gauge coupling constants, have a tree level dependence on
closed string moduli \cite{Kaplunovsky:1995jw} \footnote{Unlike the
  pseudo-moduli discussed here, which only have flat directions at
  special points in field space, these moduli have no potentials
  classically.  Non-perturbative corrections are required to give
  string moduli potentials, whereas pseudo-moduli have potentials at
  tree-level and the flat directions are removed at the one-loop
  level.} .  Since pseudo-moduli are evidently not string moduli, the
tree-level gauge kinetic function does not have to depend on them.
Hence, if the string moduli (or any fields that appear in the gauge
kinetic function) do not contribute strongly to SUSY breaking
\footnote{As we expect in certain racetrack models.  For example the
  O'KKLT class of models \cite{Kallosh:2006dv} have finely-tuned
  moduli sectors in which, when considered alone, have
  SUSY-preserving, De-Sitter minima.  The addition of an
  O'Raifeartaigh sector \cite{O'Raifeartaigh:1975pr} (the possibility
  that this might be ISS was considered in \cite{Dudas:2006gr})
  spontaneously breaks SUSY and lifts the vacuum energy.  Since the
  two sectors are decoupled at the global level the moduli F-terms can
  be significantly smaller than the other fields'.} then the gaugino
masses will receive a negligible tree-level contribution from gravity.

In summary, the models discussed here may naturally generate, via the gauge mediation channel, a spectrum in which the
gaugino masses are loop suppressed with respect to the soft scalar
masses.  Since the gaugino masses receive unknown quantum corrections,
we cannot predict the precise spectrum, which depends on details of the complete model, 
though we do expect to see a split in the spectrum of soft masses.  This
spectrum has much in common with the one presented in the early work
on anomaly mediation.  See, for example, \cite{Giudice:1998xp}.

In addition to this, one can consider modifying the model.
For example, there are examples \cite{Abel:2007jx,
  Haba:2007rj,Xu:2007az,Kitano:2006xg} in which operators are added to
the superpotential allowing the pseudo-modulus vev to grow out to
around $\mu$.  While this is clearly an interesting effect, the dual
theory responsible for generating this operator is as yet unknown.

\section{Conclusions}

In summary, ISS is more stable with supergravity corrections than
without, assuming that $W_0=0$.  The picture changes somewhat when
$W_0\neq0$, but it was shown in section \ref{sec:One Loop} that the
supergravity effects are under good theoretical control. 

It was then demonstrated, in section \ref{sec:NonPert} that, while the
non-perturbative piece is non-zero, it is necessarily sub-dominant for
small values of $\Phi$.  Hence this term can be neglected when
supergravity effects are subdominant to the one-loop effects.
Numerical analysis confirmed this.

We also showed that the quadratically divergent, one-loop potential's
effects are small and under control if $\Lambda' N_f' \ll 1$.  We have
constructed our theory such that supersymmetry breaking is only
generated by the ISS sector, to highlight the distinctive features of
this model.  An example of a possible a high energy model was
considered in \cite{Ferrara:1994kg}, but we did not attempt this kind
of construction in this paper.  Also, it is worth noting that
it is consistent to make use of the supergravity corrected effective
potential, even though our cut-off can be taken to be many orders of
magnitude below.  This is because the non-renormalisable operators
present in Sugra are not generated by the integration out of
gravitational fields, and are instead required to be present by
supersymmetry itself.

The explicit R-breaking introduced by the presence of a constant term
in the superpotential allows the generation of non-zero gaugino
masses, through gravitationally suppressed interactions (as one would
expect, since global SUSY is blind to the presence of the constant).

We showed, in section \ref{sec:SoftMasses}, that in the absence of
additional supersymmetric contributions to messenger masses, the gauge
mediation is only possible if the magnetic quarks couple to the
messenger fields.  All other ISS fields have overly small scalar vevs
and give rise to an unstable messenger sector.  The direct consequence
of which is that the gravitino mass is quite large.  This is the case
because it is not set by the quarks F-terms, but the much larger meson
F-terms, through the requirement that the cosmological constant should
vanish in the minimum of the effective potential.  This results in a
direct connection between R-symmetry breaking and gaugino masses, with
Eq.~(\ref{eq:gaugino_mass}) showing that they are proportional to one
another, with the coefficient being determined by the details of the
gauge group.

\section*{Acknowledgements}
This work was partially supported by the EC 6th Framework
Programme MRTN-CT-2006-035863, by the  grant MNiSW  N202 176 31/3844
and  by TOK Project  MTKD-CT-2005-029466.

\clearpage
\appendix

\Large \noindent {\bf Appendix}
\normalsize
\section{Origins of mass terms} \label{sec:appA}

 For simplicity we neglect phases in the
  following analysis.  Hence all symmetries effectively go from
  $U(N)\rightarrow O(N)$ and the counting of degrees of freedom
  reflects this.

In this case, the tree-level potential, when embedded into
supergravity, has a meta-stable minimum. The position of the minimum
is given by the global result, but with small corrections to the
expectation value of the magnetic quarks $|\phi_0|^2=|\tilde{\phi_0}|^2= \mu^2 -
1/2 \pm \frac12 (1 - 4 N_c \mu^4)^{1/2}\sim \mu^2-N_c \mu^4$.  This
deviation from the global limit only comes from the overall factor of
$E^K$ multiplying the potential\footnote{In the limit where $\Phi
  \rightarrow 0$ and $W\rightarrow 0$ all contributions aside from
  the overall exponential vanish.}.

The spectrum contains $2 N_f N$ magnetic quarks of which $\frac12(2N_f
N-N^2-N)$ are massless Goldstone bosons and the remainder have masses
of order $\mu$, $2N_f N - N^2$ mesons with mass of order $\mu$,
$(N_f-N)^2-1$ mesons with mass of order $\mu^2$ and one massless
pseudo-moduli meson.

The origins of these masses are as follows.  The quarks get their
masses from their expectation values, with $\phi_0$ giving mass to
$\tilde{\phi}_0$ and vice versa, the off-diagonal elements of $\Phi_0$
obtain masses solely from the second derivative of $e^K$ (which
contributes equally to all fields), namely $2 V$, while the diagonal
elements get more complicated contributions.  The remaining elements
of $\Phi$ get the same masses as in global SUSY, but with small
corrections from the Sugra contributions.  The end result of this is
that the massive fields retain essentially same masses as in global
SUSY and all but one of the pseudo-moduli (which remains zero) obtain
masses of the order of the cosmological constant: $2 N_c \mu^4$.  This
demonstrates that supergravity serves to increase the stability of the
ISS minimum, as it reinforces the stabilising effects coming from the
one loop potential.

\section{Analytical solutions} \label{sec:appB}
Here we present the approximate analytical expressions for the
derivatives of the tree-level and quadratic, one-loop potentials.  Since both potentials can be written in terms of $V_F$ and $V_W$ we compute the derivatives of these functions, taking the fields to be real.
\begin{align}
  \frac{\partial V_F}{\partial \Phi_{1}} &=2 h^2\left( \tilde{\phi}_{0}(\tilde{\phi}_0 \Phi_{1}+\phi_0 W_0) +\phi_{0}(\phi_0 \Phi_{1}+\tilde{\phi}_0 W_0)  +W_0 (\phi_0 \tilde{\phi}_0-\mu^2)\right)+\mathcal{O}(\mu^6) \sim h^2 \mu^4 \\
  \frac{\partial V_F}{\partial \Phi_0} & = - 2 h^2 W_0 \mu^2 +\mathcal{O}(\mu^6) \sim h^2 \mu^4 \\
  \frac{\partial V_F}{\partial \phi_0} & = 2 h^2 \tilde{\phi}_0 (\phi_0 \tilde{\phi}_0-\mu^2) +\mathcal{O}(\mu^5) \sim h^2 \mu^3
\end{align}

The derivatives of $V_W=-3 e^K W^2$ are easily computed and given below, again taking the fields to be real
\begin{align}
    \frac{\partial V_W}{\partial \Phi_{1}} &= -6h^2 W_0 (\tilde{\phi}_0 \phi_0-\mu^2)  +\mathcal{O}(\mu^6) \sim h^2 \mu^4 \\
  \frac{\partial V_W}{\partial \Phi_0} & = 6 h^2 W_0 \mu^2 +\mathcal{O}(\mu^6)  \sim h^2 \mu^4\\
  \frac{\partial V_W}{\partial \phi_0} & = -3 h^2 W_0\tilde{\phi}_0 \Phi_0 -3 h^2 \phi_0 W_0^2  +\mathcal{O}(\mu^7) \sim h^2 \mu^5
\end{align}

Hence we see that while both $\Phi_0$ and $\Phi_1$ depend directly on $W$, $\phi_0$ does not.  This implies that it will remain at the global SUSY minimum, up to corrections induced by the logarithmic piece.

We can make use of the global expression for the logarithmic contribution to $\Phi_0$'s potential,
\begin{align}
V_{\mbox{log}}=\frac{h^4\mu^2(\log(4)-1)}{8\pi^2}\mbox{Tr}(\Phi_0)^2 \label{eq:v_log_ISS}
\end{align}
and estimate $\Phi_0$'s expectation value, without the quadratic contribution,
\begin{align}
   \<\Phi_0\>=  \frac{16 \pi^2 W_0}{h^2(\mbox{log}(4)-1)}+ \mathcal{O} (\mu^4), \label{Eq:Phi0}
\end{align}

While we do not have estimates for the logarithmic contributions for $\Phi_1$, $\phi_0$ and $\tilde{\phi}_0$, we can obtain the tree-level expectation values,
\begin{align}
 \<\Phi_1 \>= \frac{- \mu^2 W_0}{\phi_0 \tilde{\phi_0}} +  \mathcal{O} (\mu^4) \label{Eq:Phi10}
\end{align}
and
\begin{align}
  \<\phi_0\>= \<\tilde{\phi}_0\> = \mu +  \mathcal{O} (\mu^3) \label{Eq.phi0}
\end{align}
and from this point forward we take $\phi_0= \tilde{\phi}_0$.  Eq.~(\ref{Eq.phi0}) and Eq.~(\ref{Eq:Phi10}) give
\begin{align}
 \<\Phi_1 \>= - W_0 +  \mathcal{O} (\mu^4). \label{Eq:Phi1}
\end{align}
These are the results previously obtained in the literature, under the
assumption that the one-loop potential presented in
\cite{Intriligator:2006dd} was valid away from the minimum in which it
was derived and that the potential for $\Phi_1$ and $\phi_0$ is flat.
However, we show numerically that neither of these assumptions are
valid, given $h\sim 1$, and hence the one-loop potential plays a more
significant role than has been previously discussed.

In addition to this, we can compute the effects of the quadratic, one-loop potential.  Taking Eq.~(\ref{Eq:One-loop breakdown}) and ignoring the log piece, we obtain
\begin{align}
   V_{\mbox{tree}}+ V_{\mbox{quad.}} = \left(N_f' \Lambda'^2 + 1 \right) V_{F} -  ( 3 + 2 \Lambda'^2(N_f'+1)) e^K W^2.
\end{align}

First we observe that $\<\phi_0\>$ will be unchanged as the quadratic contribution, at leading order, simply increases the coefficient of $V_F$ from 1 to $1+\Lambda'^2 N_f'$.    
\begin{align}
  \frac{\partial ( V_{\mbox{tree}}+ V_{\mbox{quad.}})}{\partial \Phi_{1}} & = 4 h^2 (N_f' \Lambda'^2 + 1) \phi_0^2 \Phi_1 + 2 h^2 \phi_0^2 W_0(N_f'-2)\Lambda'^2 + 2 h^2 \mu^2 W_0 ( 2 + 2 \Lambda'^2 + N_f' \Lambda'^2) \\
  \frac{\partial (V_{\mbox{tree}}+ V_{\mbox{quad.}} )}{\partial \Phi_0} & = 4 h^2 W_0 \mu^2 \left(1+ \frac{\Lambda'^2}{2} (N_f'+2) \right)  +\mathcal{O}(\mu^6) \sim h^2 \mu^4 
\end{align}
and hence 
\begin{align}
 \<\Phi_0 \>= \frac{ 16\pi^2 W_0 (1+\frac{\Lambda'^2}{2}(N_f'+2))}{h^2(\mbox{log}(4)-1)} +  \mathcal{O} (\mu^4) 
\end{align}
and 
\begin{align}
 \<\Phi_1 \>= \frac{- W_0 (\phi_0^2 \Lambda'^2(N_f'-2 ) +\mu^2(2+2 \Lambda'^2 + N_f' \Lambda'^2)}{2\phi_0^2 (N_f' \Lambda'^2+1)} +  \mathcal{O} (\mu^4)= - W_0  + \mathcal{O} (\mu^4).
\end{align}

It is interesting to note that the quadratic potential reinforces
the tree-level solution, since, when $\phi_0\rightarrow \mu$, they have the same form.  We confirm this in
section~\ref{sec:Numerics}.

\end{document}